\documentclass[11pt]{article}

\usepackage[margin=1in]{geometry}
\usepackage{amsmath}
\usepackage{amsfonts}
\usepackage{array}
\usepackage{booktabs}
\usepackage{graphicx}
\usepackage{hyperref}
\usepackage{url}
\usepackage{float}
\usepackage[section]{placeins}
\usepackage[numbers,sort&compress]{natbib}

\graphicspath{{./}}

\newcommand{\affmark}[1]{\raisebox{0.6ex}{#1}}

\title{From Clerks to Agentic AI: How Will Technology Transform the Labor Market in Finance?}
\author{Lu Yu\affmark{1} and Xiang Li\affmark{2}}
\date{}

\makeatletter
\renewcommand{\maketitle}{%
  \begingroup
  \renewcommand{\thefootnote}{\fnsymbol{footnote}}%
  \newpage
  \null
  \vskip 2em%
  \begin{center}%
    {\LARGE \@title \par}%
    \vskip 1.5em%
    {\large
      \lineskip .5em%
      \begin{tabular}[t]{c}%
        \@author
      \end{tabular}\par}%
    \vskip 1em%
    {\small \affmark{1} Department of Economics, Georgetown University \par}
    {\small \affmark{2} Pittsburgh Supercomputing Center, Carnegie Mellon University \par}
    {\small Equal contribution; correspondence: ly229@georgetown.edu \par}
  \end{center}%
  \par
  \vskip 1.5em%
  \endgroup
}
\makeatother

\begin{document}
\maketitle

\begin{abstract}
Artificial intelligence is beginning to reshape financial markets in ways that echo earlier technological transitions in finance. This paper examines how task automation changes labor productivity and labor demand in U.S. financial firms across three technology waves: computerization, indexing, and agentic AI. Combining historical comparison with firm-level operating outcomes and filing-based automation measures, it argues that successive automation waves reorganize workflows, scale, and the distribution of capability across firms rather than simply eliminating jobs. The resulting pattern is likely to be a more polarized industry in which large institutions retain structural advantages, small teams become more capable, and middle-layer functions face the greatest pressure.
\end{abstract}

\section{I. How Does Agentic AI Impact Financial Markets?}
Finance is an unusually informative setting for studying automation because it combines standardized workflows, information processing, client service, and judgment-intensive decision making within the same firms. New technology therefore affects tasks unevenly: some activities become cheaper and faster almost immediately, while others remain constrained by supervision, trust, interpretation, and accountability. The relevant question is not simply whether AI reduces headcount, but how automation changes the internal allocation of tasks and what that implies for productivity, scale, and labor demand.

The analysis links historical technological change to contemporary AI adoption. Rather than treating AI as a purely speculative future shock, it examines whether rising automation intensity in firm disclosures is associated with changes in revenue per employee, assets under management per employee, and labor expense per employee. The broader aim is to place agentic AI within a longer sequence of technological change in finance rather than to treat it as a complete break from the past.

The contribution is threefold. First, the paper organizes the analysis around three economically meaningful waves: computerization, indexing, and AI-driven automation. Second, it constructs filing-based measures of automation intensity that map onto these waves. Third, it relates those measures to firm-level operating outcomes to assess whether finance becomes more scalable and productive before labor-cost adjustment becomes visible.

The preliminary evidence is consistent with that view. Productivity rises across successive technology waves, labor-cost adjustment appears slower than output adjustment, and AI-era firms exhibit the strongest filing-based automation intensity. These patterns are descriptive associations, not causal estimates, but they provide a disciplined framework for studying how automation may reorganize labor in finance.

\section{II. Historical Lessons from the Computerization and Indexing Eras}
The natural benchmark for the current AI wave is the earlier computer revolution in finance. In the 1980s and 1990s, personal computers, spreadsheets, Bloomberg terminals, Reuters systems, and electronic trading infrastructure transformed how financial firms processed information and managed risk. Those technologies did not eliminate finance; they changed who could operate effectively, how quickly information moved through the system, and which firms could scale their edge.

The second major wave came with indexing and rules-based investing. Between roughly 2000 and 2015, ETFs, benchmark replication, passive allocation, and automated portfolio implementation reshaped large parts of asset management and market intermediation. This wave mattered not because it replaced all active judgment, but because it codified a large set of investment and rebalancing tasks that had previously been more labor intensive or discretionary.

The third wave is the current AI and workflow-automation era, beginning around 2015 and accelerating sharply with generative and agentic systems. Unlike the earlier computerization wave, which mainly expanded computation and connectivity, the AI wave pushes further into cognition-adjacent tasks such as monitoring, summarization, drafting, triage, and workflow coordination. Unlike indexing, which standardized portfolio construction in a narrower domain, AI has the potential to affect a broader range of occupations inside financial firms.

These eras are descriptive and interpretive rather than a stand-alone identification strategy. Their purpose is to provide a historically grounded framework for comparing how different forms of task automation alter productivity and labor demand over time.

The task-level economic logic is illustrated by the following cost comparison, which shifts the discussion from infrastructure to the practical choice among domestic labor, offshore labor, and AI agents for repeatable workflows.

\begin{figure}[H]
  \centering
  \includegraphics[width=\linewidth]{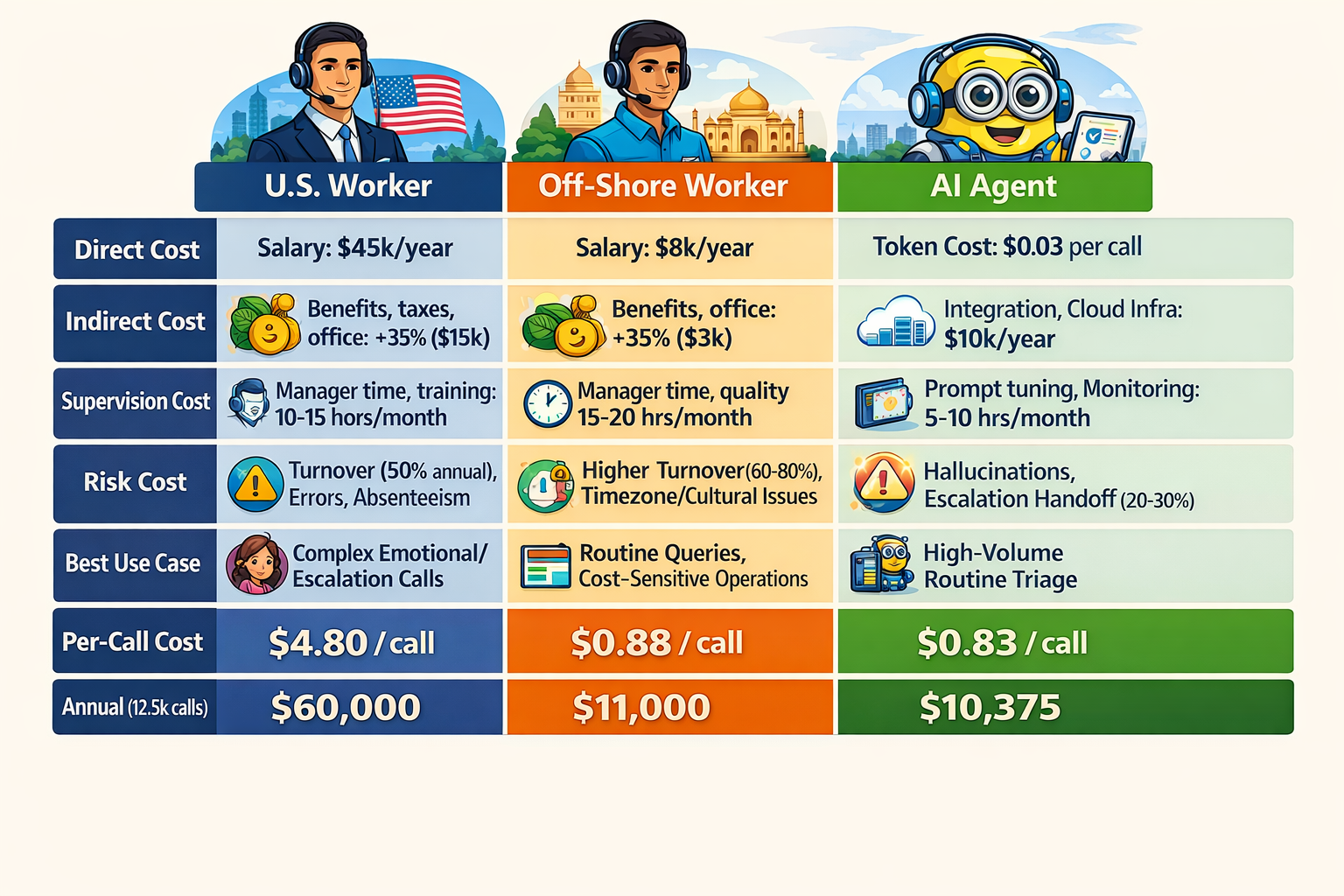}
  \caption{Workforce transformation cost comparison}
\end{figure}

\section{III. How Does Automation Raise Labor Productivity?}
Automation can raise labor productivity through several channels: it can reduce routine manual work, lower coordination costs, accelerate workflow speed, and enable firms to operate at greater scale without proportionate increases in labor input. In finance, these effects may appear in back-office processing, reporting, risk monitoring, trade support, and parts of research production.

Automation, however, need not generate immediate labor displacement. Technology can change the composition of work well before it changes headcount. Firms may reassign workers toward supervision, exception handling, and client-facing tasks while software absorbs repetitive tasks in the background. Labor expense per employee may therefore respond more slowly than output-based measures of productivity.

The core conceptual variable is \texttt{task\_automation\_intensity}, defined as the extent to which routine, codifiable, or repetitive work is shifted from human labor to software, rules, machines, or AI systems. The vocabulary changes across eras, but the economic concept remains stable: firms differ in how intensively they automate tasks previously performed by people.

This framework yields three testable implications. First, productivity should rise more in later technology eras than in the pre-computerization benchmark. Second, higher task-automation intensity should be associated with higher output per worker or greater scale per worker. Third, labor expense per worker may move more slowly than output measures if automation primarily works through reorganization and augmentation rather than immediate labor shedding.

\section{IV. How Do We Define Productivity and Automation Intensity?}
\subsection{A. Who Do We Study?}
The empirical setting is a panel of publicly listed U.S. financial firms observed in filing and accounting data. The broader design targets a sample of roughly 100 firms, with a possible extension to 300 firms for robustness or appendix analysis. In the current draft, the AI-specific filing analysis uses a focused sample of five large institutions: Bank of America, BNY Mellon, JPMorgan Chase, S\&P Global, and State Street.

Firm-years are assigned to the three broad technology eras using calendar-year cutoffs: \texttt{Computerization} (\texttt{1985-2000}), \texttt{Index investing} (\texttt{2000-2015}), and \texttt{AI / automation} (\texttt{2015-present}). These cutoffs are stylized, but they provide a transparent way to compare broad phases of technological change in finance.

\subsection{B. What Is the Productivity Measure?}
The main productivity outcome is revenue per employee, which captures broad operating productivity and is available across a wide set of firms. Secondary outcomes include labor expense per employee and, where relevant, assets under management per employee. The AUM-based measure is especially informative for asset managers, custodians, and related firms in which scale can expand without a proportional increase in headcount.

Together, these outcomes distinguish different channels through which automation may matter. Revenue per employee captures overall operating productivity. AUM per employee captures scale in information-processing and oversight-intensive businesses. Labor expense per employee provides a direct, though incomplete, proxy for whether organizational automation is translating into near-term labor-cost compression.

\subsection{C. How Is Task Automation Intensity Measured Across Eras?}
The era-specific automation measures are built from annual reports and 10-K filings. For the computerization era, the relevant language includes terms related to computers, terminals, spreadsheets, electronic trading, digitization, workflow systems, and back-office automation. For the indexing era, the relevant vocabulary includes index funds, ETFs, passive management, benchmark replication, systematic rebalancing, and rules-based allocation. For the AI era, the measure includes terms such as artificial intelligence, machine learning, generative AI, autonomous agents, robotic process automation, workflow automation, and large language models.

In the current AI-focused panel, the exposure measure counts direct AI language such as ``artificial intelligence,'' ``machine learning,'' ``generative AI,'' and ``autonomous agent,'' together with adjacent automation language such as ``automation,'' ``robotic process automation,'' and ``workflow automation.'' Direct AI terms receive higher weight, and the weighted count is normalized by total filing words to form the final \texttt{ai\_exposure} variable.

More generally, each era-specific score is normalized by filing length, and the scores can be standardized within sample for comparability. If useful for the broader paper, the three measures can also be combined into one composite \texttt{task\_automation\_intensity} index.

\section{V. Our Empirical Framework}
The empirical design begins with descriptive comparisons across eras and then moves to a fixed-effects panel specification. The goal is not to claim clean causal identification, but to test whether within-firm increases in automation intensity align with systematic changes in productivity and labor-cost outcomes.

The baseline AI-era specification is:

\begin{equation}
\log(1 + \text{outcome}_{it}) = \beta \cdot \text{ai\_exposure}_{it} + \text{firm FE} + \text{year FE} + \text{error}_{it}.
\end{equation}

This design absorbs time-invariant firm characteristics and common macro shocks. It is therefore appropriate for descriptive within-firm comparisons, but it does not support strong causal interpretation. In the broader three-era framework, analogous specifications can regress productivity outcomes on era indicators, automation-intensity measures, and their interactions.

The framework also allows for heterogeneity analysis. For example, one can compare asset managers with banks, passive-heavy firms with active-heavy firms, and large institutions with mid-sized firms. These comparisons are important because automation may affect scale, coordination, and labor demand differently across business models.

\section{VI. How Does Automation Shape Productivity?}
\subsection{A. Productivity Trends Across Eras}
The earlier \texttt{simple\_analysis} provides the descriptive bridge from the computer era to the current AI period. It compares firm-group productivity across four broad eras: pre-computerization, computerization, indexing, and AI. The regression evidence shows that both \texttt{revenue\_per\_employee} and \texttt{aum\_per\_employee} are substantially higher in later eras than in the pre-computerization benchmark, even after controlling for firm-group fixed effects. In the revenue specification, the era coefficients rise from 1.4070 in the computerization period to 2.0543 in the indexing period and 2.3688 in the AI period. In the AUM specification, the coefficients rise from 1.3638 to 2.4197 and then 3.3944.

These coefficients should be interpreted as descriptive rather than causal. They nonetheless indicate that the financial panel moves in the same broad direction as the later SEC-based analysis and suggest that productivity growth in finance accumulated through successive waves of technological and organizational change rather than arriving all at once.

\begin{figure}[H]
  \centering
  \includegraphics[width=\linewidth]{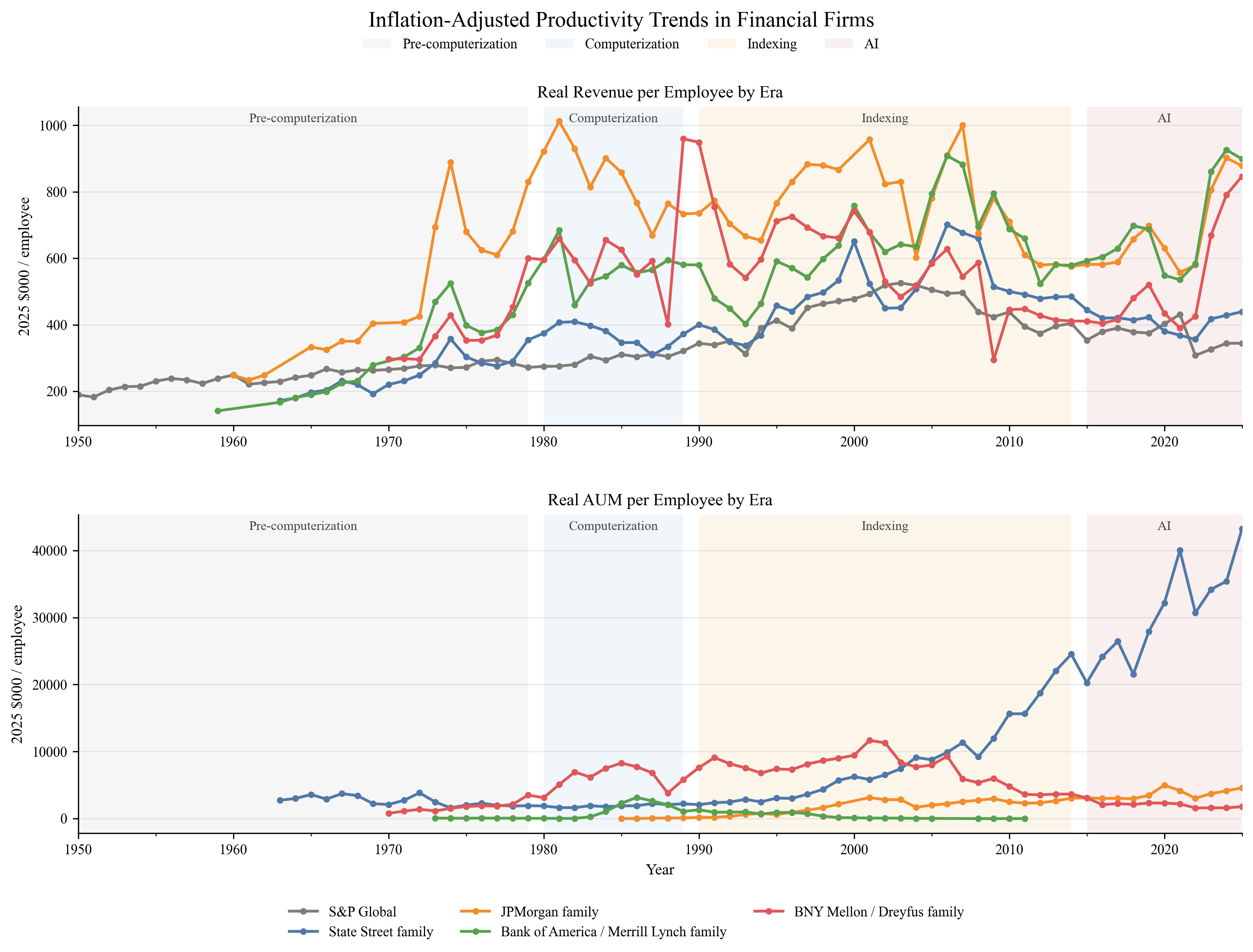}
  \caption{Simple finance productivity eras real}
\end{figure}

The real-era figure reinforces this pattern after deflating the productivity ratios, indicating that the increase is not simply an artifact of nominal growth.

\subsection{B. Automation and Productivity}
The AI-focused filing analysis examines whether firms with rising AI disclosure intensity also exhibit systematic changes in operating outcomes. The fixed-effects estimates indicate a mixed but informative pattern. The coefficient on AI exposure is negative for \texttt{revenue\_per\_employee}, at \texttt{-0.0535} with a robust standard error of \texttt{0.0117}, and positive for \texttt{aum\_per\_employee}, at \texttt{0.5843} with a robust standard error of \texttt{0.1646}.

These coefficients should be interpreted as descriptive associations rather than causal effects. Substantively, greater AI disclosure intensity is associated with lower revenue intensity per worker but higher AUM per worker. In a financial setting, that combination is more consistent with reorganization, reinvestment, and output-mix change than with a single-margin productivity story. The coefficient plot below summarizes the firm- and year-fixed-effects estimates across outcomes.

\begin{figure}[H]
  \centering
  \includegraphics[width=\linewidth]{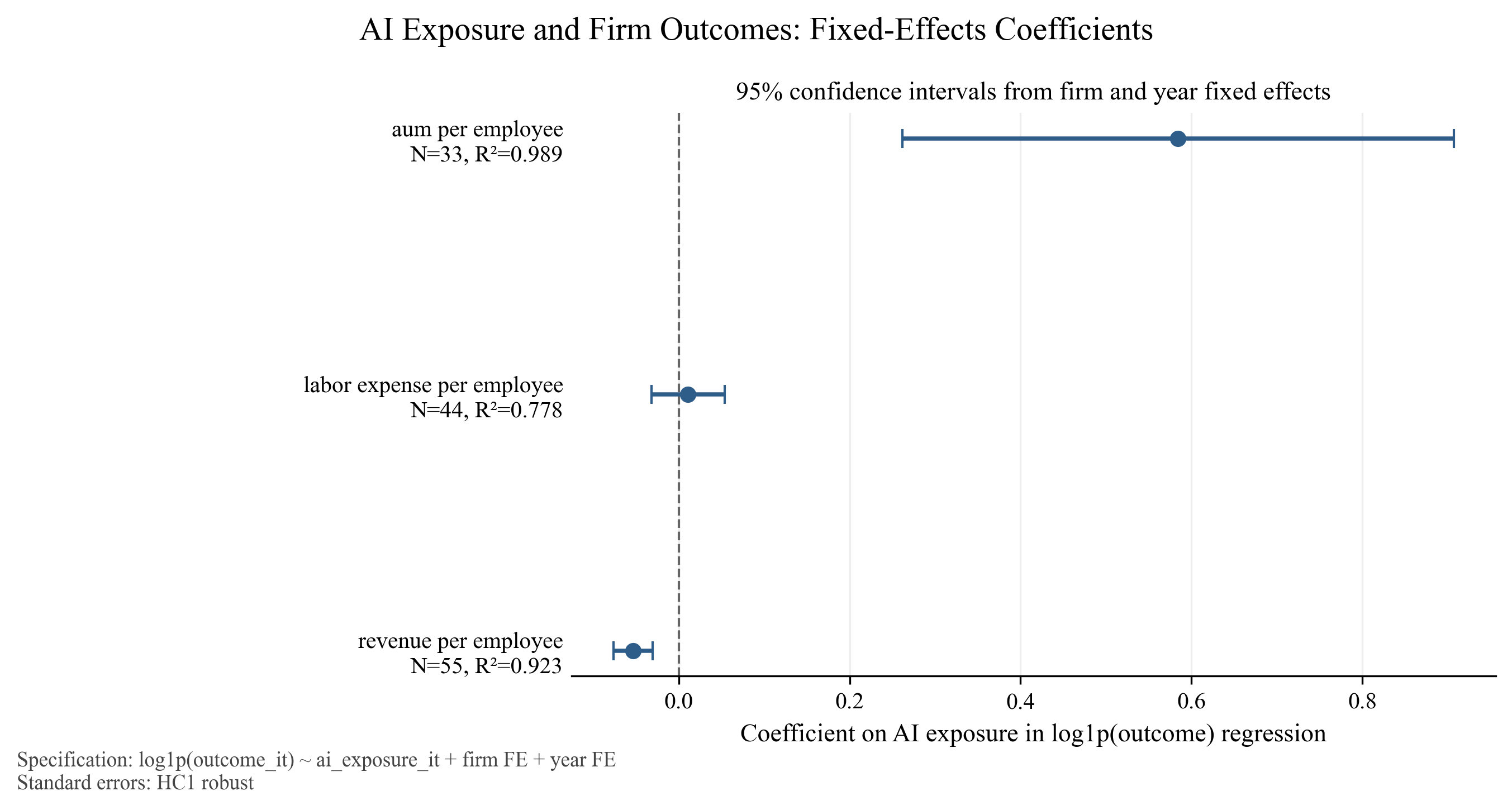}
  \caption{Coefficient plot of AI exposure fixed-effects results}
\end{figure}

\subsection{C. Labor Cost Response}
The coefficient on \texttt{labor\_expense\_per\_employee} is small and statistically indistinguishable from zero, at \texttt{0.0107} with a robust standard error of \texttt{0.0219}. This pattern suggests that firms do not respond to rising automation intensity through immediate, visible labor-cost compression. Instead, AI may initially affect scale, task allocation, and the organization of work before those changes appear in compensation or headcount measures.

The wage-trend evidence from the earlier panel is consistent with that interpretation. Labor costs evolve over time, but not in lockstep with output-per-worker measures, as would be expected if automation primarily operates through augmentation, workflow consolidation, and task reallocation.

\begin{figure}[H]
  \centering
  \includegraphics[width=\linewidth]{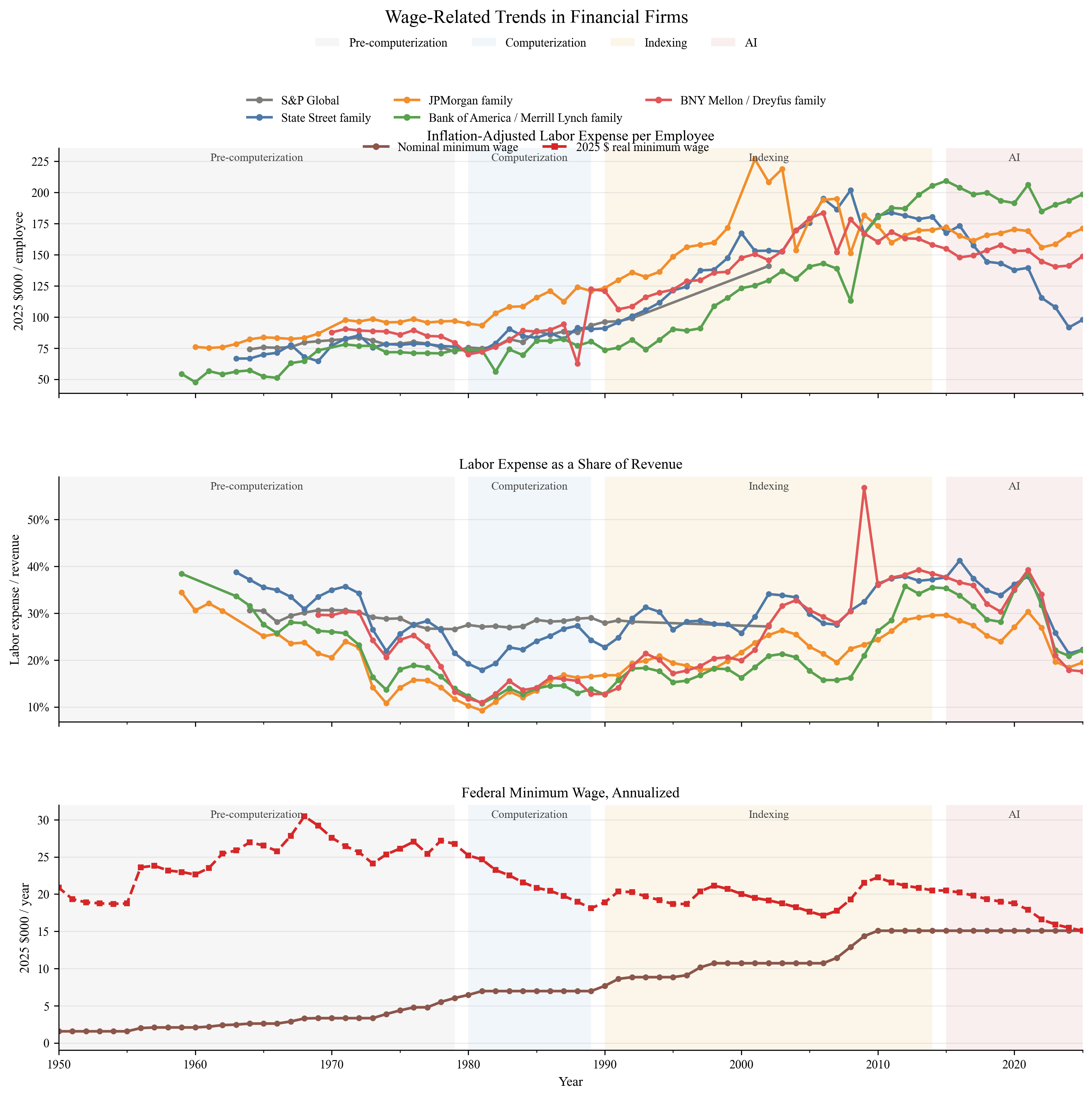}
  \caption{Wage trends by group real}
\end{figure}

\subsection{D. Heterogeneity}
The event-study style figure suggests that adoption is staggered across firms rather than synchronized at the industry level. Bank of America appears earlier, S\&P Global accelerates later, JPMorgan and State Street move more gradually, and BNY Mellon crosses the threshold very late in the sample. This timing pattern points to heterogeneous organizational adoption rather than a single sector-wide break. The figure aligns each firm's exposure series around its first substantive increase in disclosure intensity.

This heterogeneity is economically important. Large institutions retain advantages in infrastructure, proprietary data, compliance depth, and execution quality. By contrast, firms whose historical edge depended on labor-heavy coordination may face greater pressure if AI lowers the cost of standardizing information flows and support functions. Extending the analysis to compare banks with asset managers, passive-heavy with active-heavy firms, and large with mid-sized institutions is therefore a natural next step.

\begin{figure}[H]
  \centering
  \includegraphics[width=\linewidth]{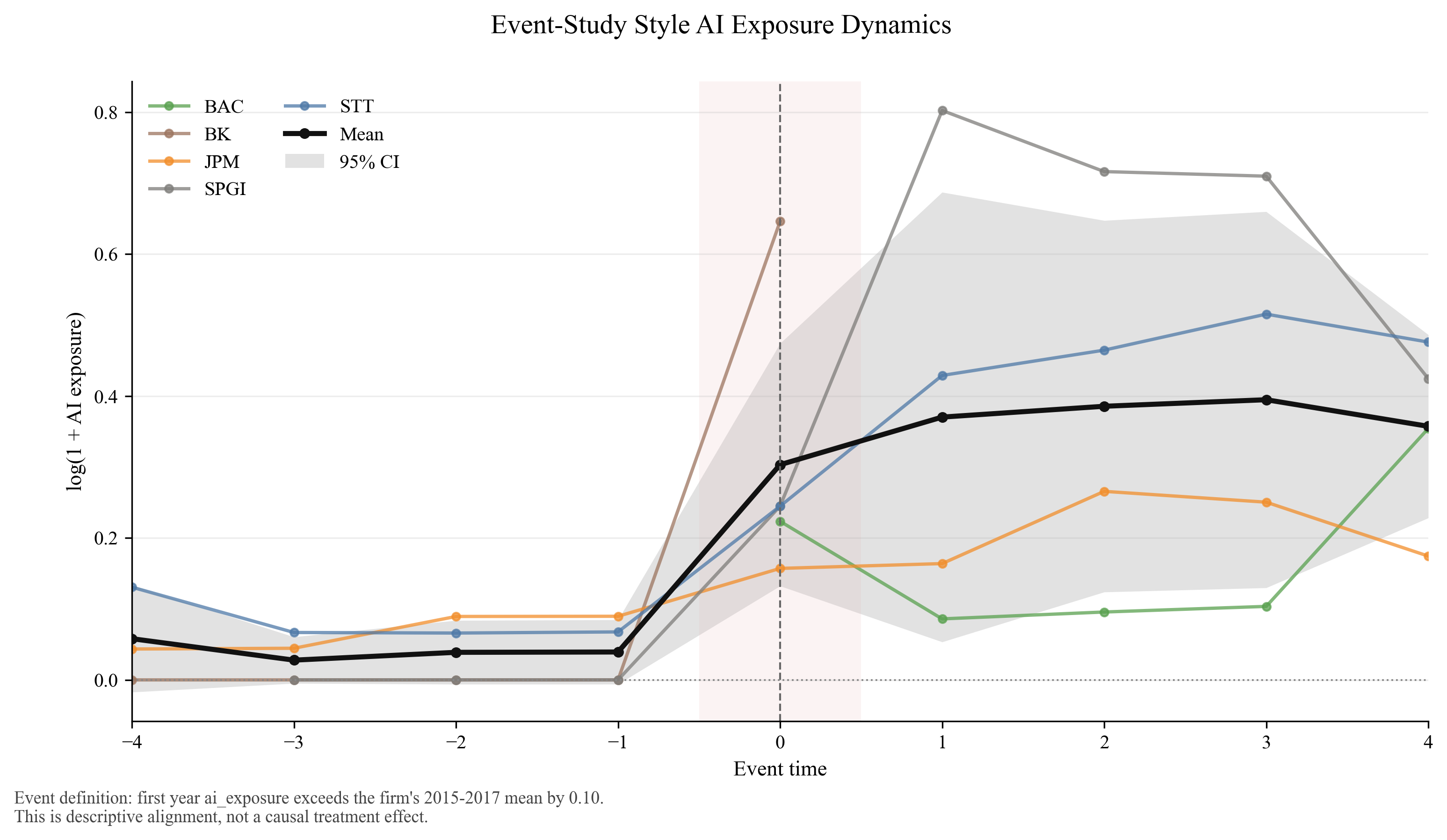}
  \caption{Event-study style AI exposure dynamics}
\end{figure}

\section{VII. Transformation for the Financial Labor Market}
Taken together, the evidence points to transformation rather than simple replacement. Productivity gains appear before labor-cost reductions, suggesting that automation in finance initially operates by changing how work is organized rather than by immediately eliminating jobs. Firms can use technology to expand monitoring capacity, standardize routine workflows, and scale output per worker while still relying on human labor for oversight, interpretation, client trust, and strategic judgment.

This logic also helps explain why indexing and AI both matter, but in different ways. Indexing codified and standardized a narrower set of portfolio and execution tasks, especially in asset management. AI reaches more deeply into cross-functional workflows such as monitoring, reporting, drafting, triage, and coordination. The implication is that AI can affect a broader set of occupations even if the immediate accounting response looks gradual.

One likely outcome is a more polarized industry. Large institutions preserve structural advantages in data, infrastructure, and compliance. Small teams become more capable because AI compresses the cost of research support, market monitoring, and operational coverage. The middle layer faces the greatest pressure because its historical value often came from coordinating information flows that increasingly can be standardized or partially automated.

This interpretation is consistent with the broader automation literature. Autor's task-based framework emphasizes that technology substitutes for routine tasks while complementing abstract and interpersonal ones. Acemoglu and Restrepo distinguish automation, augmentation, capital deepening, and task creation as separate channels. Brynjolfsson and McAfee emphasize that organizational redesign matters as much as the technology itself. The results here fit that logic: automation in finance appears to align more clearly with reorganization of scale and workflow than with an immediate decline in labor expense per employee.

\section{VIII. What Are the Study's Limitations?}
This analysis remains exploratory and has several limitations. First, the filing-based exposure measure captures disclosure intensity as well as underlying operational adoption. Firms may differ in how much they discuss AI even when actual implementation differs less. Second, the current AI-focused sample includes only five firms, so the results should be viewed as a structured pilot rather than a definitive sector-wide estimate.

Third, the fixed-effects design is descriptive and cannot cleanly separate causality from reverse causality or omitted organizational change. Productivity ratios are also influenced by business mix, firm strategy, and market conditions. Finally, the three-era classification is stylized. It is useful as a historical framework, but it should not be interpreted as a sharp causal partition of technological change.

\section{IX. Conclusion}
This paper develops a framework for studying how agentic AI changes the labor market in finance by embedding the current AI wave within a longer history of computerization and indexing. Its main contribution is a transparent automation-intensity approach built from annual filings and linked to firm-level operating outcomes. Across the three-era perspective, finance appears to become more scalable and more productive over time, but labor adjustment remains uneven and gradual.

The current AI evidence indicates that higher AI exposure is associated with greater AUM per employee, weaker revenue-per-employee performance, and no clear immediate change in labor expense per employee. The event-study evidence likewise suggests that adoption is staggered and organizationally mediated. More broadly, the results suggest that technology is likely to transform the labor market in finance through task reallocation, workflow redesign, and uneven gains in capability rather than through uniform labor replacement.

\end{document}